\documentclass[%
 amsmath,amssymb,
 aps,nofootinbib,
 twocolumn,
]{revtex4-1}
\usepackage[usenames, dvipsnames]{color}
\usepackage[autostyle]{csquotes}
\usepackage{pgfplots}
\usepackage{tikz}
\usepackage{graphicx}% Include figure files
\usepackage{dcolumn}% Align table columns on decimal point
\usepackage{bm}% bold math
\usepackage{enumerate}% bold math
\usepackage{setspace}
\usepackage{float}
\usepackage{hyperref}
\definecolor{MyDarkBlue}{rgb}{0,0.1,0.7}
\hypersetup{colorlinks,breaklinks=true,
  urlcolor={blue},citecolor={MyDarkBlue},linkcolor={Blue}}
\begin{document}
\title{Isocurvature modes and non-Gaussianity in affine inflation}% Force line breaks with \\
\author{Hemza Azri$^{1}$}
\email{hmazri@uaeu.ac.ae; hemza.azri@cern.ch}%Lines break automatically or can be forced with \\
\author{Isaac Bamwidhi$^{1}$}%
\author{Salah Nasri$^{1,2}$}%
\email{snasri@uaeu.ac.ae; salah.nasri@cern.ch}
\affiliation{$^{1}$%
 Department of Physics, United Arab Emirates University,
Al Ain 15551 Abu Dhabi, UAE
}%
\affiliation{$^{2}$% 
International Center for Theoretical Physics, Trieste, Italy}%

\date{\today}% It is always \today, today,

\begin{abstract}
Inflationary dynamics driven by multiple fields, especially with nonminimal couplings, allow for highly interesting features such as isocurvature, non-Gaussianity, and preheating. In this paper, we study two-field inflation in the context of purely affine gravity, where the metrical structure results from the dynamics of the spacetime affine connection. We introduce a covariant formulation in the new framework and show that it leads to a curved field space which can produce conspicuous departure from the purely metric gravity. In the case where the fields are canonical, the field manifold gains a conformally flat shape. Interestingly, while the manifold is generally curved, it is possible to be flattened by allowing a specific non-canonical field kinetic terms. This in turn simplifies the inflationary dynamics significantly while allowing for new predictions due to the effects of the nonminimal coupling function on the potential solely. We use this new feature in studying two-field inflation driven by quartic potentials for a given parameter constants. We perform a numerical solution of the slow-roll dynamics and track the possible non-Gaussianity. We focus on field parameters that allow for spectral indices within the favored region of the Planck results. These are associated to tiny tensor-to-scalar ratios, $r \sim 10^{-6}$ for single field and $r \sim 10^{-4}$ for two-fields. We also show that two-field Higgs inflation may favor a curved field manifold.    
\end{abstract}

\maketitle

\section{Introductory remarks and motivations}

%The enormous progress made in high energy physics and cosmology has offered novel theoretical ideas for understanding the physics of the early universe that has been a challenging endeavour for decades \cite{early_universe}. A paradigmatic proposed idea is inflation; a scenario with remarkable predictions that become accurately testable  by the current cosmological observations \cite{inflation} (see also \cite{martin} for a modern perspective). Standard inflation stands on general relativity (GR) as a theory of gravity and posits an early phase of rapid expansion to be driven by scalar fields. While it offers  possible solutions to the shortcomings of the hot big bang model, it incredibly serves as a relevant mechanism for the origin of structure.

Strong supports of the inflationary models are mainly their predictions of the power spectrum of the primordial perturbations that have been relatively in agreement with the observations of the Cosmic Microwave Background (CMB) anisotropies carried out by the most recent Planck results \cite{planck}. Nevertheless, the existence of numerous models of inflation that fit the current data, suggests that a more credible inflationary paradigm needs to be determined. Moreover, there has been a lot effort  to  reconcile a plausible inflationary paradigm with realistic models of elementary particles or with well-motivated models of high energy physics \cite{hep_inspired}. In this respect, detailed studies have been performed on inflation with multifields coupled nonminimally to gravity \cite{kaiser_covariant, kaiser, sasaki}.

An interesting and generic feature of the multiple field models is the production of non-adiabatic (isocurvature) perturbations that could survive on superhorizon scales \cite{multi1, multi2, multi3, multi4}, as well as the non-Gaussian distribution of the perturbations. Unlike the case of single field models where these isocurvature modes are completely suppressed in the long wavelength limits, in multiple field models, they can in principle amplify the curvature perturbations and alter its evolution even after they have crossed outside the horizon. Now the fact remains whether this isocurvature and non-Gaussianity could be tracked through the temperature anisotropies of the cosmic microwave background. It turned out that a less power in the angular power spectrum of temperature anisotropies in the CMB at low multipoles, has been observed, compared to best-fit $\Lambda$CDM cosmology.  Since the cosmological perturbation in standard cosmology is known to be adiabatic and nearly Gaussian, then one might consider also the possibility of isocurvature modes (or non-Gaussianity) to account for this deviation \cite{kaiser_after_planck}. 

These features and others have been extensively studied in metrical theories of gravity, such as general relativity and its possible extensions \cite{kaiser_covariant, kaiser, sasaki, multi1, multi2, multi3, multi4, kaiser_after_planck}, as well as Palatini inflation \cite{palatini_version}. However, since it is highly possible that the inflationary dynamics depends on the assumed underlying theory of gravity, it is natural then to consider another quite different possibility where gravity is, rather, purely affine. Recently, affine gravitation with scalar fields has been applied to various phenomena \cite{scalar_connection, affine_dm, inducing_gravity, separate_einstein_eddington}.  

In a previous work, multiple scalar fields coupled to gravity through spacetime connection solely is thoroughly studied in the case of nonminimal coupling \cite{entropy_production_in_affine_inflation}. A remarkable feature drawn from this study is that both nonminimally and minimally coupled scalar fields are accommodated by a unique spacetime metric unlike the familiar theories in which the transformation to minimal couplings (Einstein frame) necessitates a second metric  obtained from the original one (Jordan frame) by Weyl transformations. This alleviate the famous frame-issue since most of the important cosmological parameters such as  the Hubble parameter, number of e-foldings,  and the gauge invariant curvature perturbation, which are highly related to the proposed metric, are not altered by the transition to minimal coupling if no conformal transformation is applied. In this respect, one of the physical consequences of this feature is that the notion of adiabaticity (of the perturbations) is frame-independent \cite{entropy_production_in_affine_inflation}.

In the present work we will be interested in the transition to minimal coupling and develop a covariant formalism of the inflationary dynamics in the context of purely affine gravity for the first time. We study two important cases that cannot be realized in purely metric gravity (GR with nonminimal coupling). The first case is that, when the fields are canonical in the beginning, the induced field space metric from the affine gravity tends to be conformal to flat. For the second case study, we will show that the field space can be made Euclidean if one sets the same coupling function in both spacetime curvature and field kinetic terms, i.e, $f(\bm{\phi})R_{\mu\nu}(\Gamma)$ and $f(\bm{\phi})\bm{\delta}_{ab}\nabla_{\mu}\bm{\phi}^{a}\nabla_{\mu}\bm{\phi}^{b}$ respectively. In this case, the fields become easily canonical but the potential gains a new structure, and this  simplifies the inflationary dynamics significantly while gaining new predictions due to the effects of the nonminimal coupling function on the potential solely. We analyse the single field-limit of this case for a quartic potential and derive the observed quantities such as the spectral index and the amount of gravitational waves. The latter is characterized by a small tensor-to-scalar ratio of the order $r \sim 10^{-6}$ for a relatively strong curvature-coupling.

We analyze the background evolution of the two-field inflation in our framework by assuming a potential in powers of the fields up to quartic terms. We show how the faltness of the field space simplifies the slow-roll dynamics and we solve numerically for the background equations and the power spectrum of the perturbation. We show that the chosen parameter constants which lead to a scalar spectral index ($n_{s}$) within the Planck region produces in turn a ratio $r \sim 10^{-4}$. We also focus on the numerical results of the three-point correlation function and the behaviour of the reduced bi-spectrum $f_{NL}$ in terms of the amount of inflation. While $f_{NL}$ vanishes for large e-folds $N$, the results show that the distribution can be less Gaussian for small $N$.

The rest of the paper is organized as follows: in section \ref{sec:affine_formulation}, we present an overview of multiple fields coupled to gravity in the purely affine formulation where the nonminimal interactions appear explicitly in the invariant action. We highlight the new features by focusing on the minimal coupling case where we use the covariant formulation. We study two important shapes of the field manifold, namely the conformally flat and the flat field spaces. Then, we analyse the inflationary dynamics in the latter case and present our predictions for quartic potentials. We conclude the paper in section \ref{sec:conclusion}. 
 
\section{ Multifield inflation: Affine gravity approach}
\label{sec:affine_formulation}
\subsection{Nonminimal couplings and nonadiabatic perturbations}
In this section we study in details  the dynamics multiple scalar fields in affine gravity and its new features when applied to inflation.

The purely affine formulation is based on the fact that the curvature, which is the essence of the geometric description of the gravitational phenomenon, is tightly related to the spacetime connection. The latter provides us with the rule for parallel displacements and defines geodesics for freely falling bodies, throughout the curved background. In this formulation, the theory must arise from only an affine connection and its curvature with no \textit{prior} notion for metric. Here the spacetime is endowed with a symmetric connection from which one forms the symmetric part of the curvature $\bm{R}_{\nu\mu}(\Gamma)=\bm{R}_{\mu\nu}(\Gamma)$. Multiple scalar fields $\bm{\phi}^{a}(x)$ with $a =1, \dots, N$, and $N$ being the number of fields, are coupled to gravity through the metric-free invariant action \cite{entropy_production_in_affine_inflation} 
\begin{eqnarray}
\label{action1}
S[\Gamma, \bm{\phi}]=
\int d^{4}x \frac{\sqrt{ |f(\bm{\phi})\bm{R}_{\mu\nu}(\Gamma)- \nabla_{\mu}\bm{\phi}^{a} \nabla_{\nu}\bm{\phi}^{a}|}} 
{V(\bm{\phi})} \nonumber \\
\end{eqnarray}
where $V(\phi)$ is a nonzero potential energy. Varying the action  with respect to the connection leads to the gravitational field equations
\begin{eqnarray}
\label{dynamical_equation}
\nabla_{\lambda} \left(f(\bm{\phi}) \frac{\sqrt{|K(\Gamma, \bm{\phi})|}}{V(\bm{\phi})}(K^{-1})^{\mu\nu} \right)=0,
\end{eqnarray}
where we have defined the tensor
\begin{eqnarray}
K_{\mu\nu}(\Gamma, \bm{\phi})=
f(\bm{\phi})\bm{R}_{\mu\nu}(\Gamma)- \nabla_{\mu}\bm{\phi}^{a} \nabla_{\nu}\bm{\phi}^{a}.
\end{eqnarray}
The dynamical equation (\ref{dynamical_equation}) is solved in terms of a generated metric tensor $g_{\mu\nu}$ satisfying
\begin{eqnarray}
\label{metric}
f(\bm{\phi}) \frac{\sqrt{|K(\Gamma, \bm{\phi})|}}{V(\bm{\phi})}(K^{-1})^{\mu\nu}=
M^{2}_{\text{Pl}}\sqrt{|g|}(g^{-1})^{\mu\nu}
\end{eqnarray}
This identity is equivalent to the following compact form of the gravitational equation
\begin{eqnarray}
\label{equations_in_terms_of_ricci}
&&K_{\mu\nu}(g, \bm{\phi})=\frac{M^{2}_{\text{Pl}} V(\bm{\phi})}{f(\bm{\phi})}  g_{\mu\nu}, \\ 
&&\nabla_{\lambda} g_{\mu\nu}=0,
\label{compatibility}
\end{eqnarray}
where the connection is reduced to the Levi-Civita connection of the generated metric thanks to identity (\ref{compatibility}).

Below we restrict the study to a two-dimensional field space and denote the fields by $\bm{\phi}^{a}=(\phi, \chi)$.
In terms of Einstein tensor, the previous equations can be recast in the following familiar form
\begin{eqnarray}
\label{eom-gravity}
R_{\mu\nu}-\frac{1}{2}g_{\mu\nu}R=
\frac{1}{f(\phi, \chi)} &&\Big\{
\nabla_{\mu}\phi \nabla_{\nu}\phi -\frac{1}{2}\nabla^{\lambda}\phi \nabla_{\lambda}\phi g_{\mu\nu} \nonumber \\
&&- \nabla_{\mu}\chi \nabla_{\nu}\chi -\frac{1}{2}\nabla^{\lambda}\chi \nabla_{\lambda}\chi g_{\mu\nu} \nonumber \\
&&-\frac{M^{2}_{\text{Pl}} V(\phi, \chi)}{f(\phi, \chi)} g_{\mu\nu}\Big\}.
\end{eqnarray}

%The second set of equations are derived by variation with respect to the fields $\phi$ and $\chi$ which yields
%\begin{eqnarray}
%\label{eom-field dynamics}
%\Box \phi -\frac{\partial V}{\partial \phi} +\frac{1}{2} \left(\frac{\partial f}{\partial \phi}\right)R =&&
%\frac{\partial \ln f}{\partial \phi} \left(\nabla^{\lambda}\phi \nabla_{\lambda}\phi + %\nabla^{\lambda}\chi \nabla_{\lambda}\chi \right) \nonumber \\
%&&+\left(\frac{M^{2}_{\text{Pl}}}{f} -1 \right)\frac{\partial V}{\partial \phi},
%\end{eqnarray}
%and
%\begin{eqnarray}
%\label{eom-field dynamics}
%\Box \chi -\frac{\partial V}{\partial \chi} +\frac{1}{2} \left(\frac{\partial f}{\partial \chi}\right) R =&&
%\frac{\partial \ln f}{\partial \chi} \left(\nabla^{\lambda}\phi \nabla_{\lambda}\phi + \nabla^{\lambda}\chi \nabla_{\lambda}\chi \right) \nonumber \\
%&&+\left(\frac{M^{2}_{\text{Pl}}}{f} -1 \right)\frac{\partial V}{\partial \chi}.
%\end{eqnarray}
%where we have used expression (\ref{metric}) to get the familiar metrical form of the %equations.
%

These equations will be adapted now to the flat FRW spacetime. The time-evolution of the background fields $\phi \sim \phi(t)$ and $\chi \sim \chi(t)$ are obtained from the gravitational equations (\ref{eom-gravity}) and written as
\begin{eqnarray}
\label{hubble_parameter}
3H^{2}=
\frac{1}{f(\phi, \chi)}
\left( \frac{\dot{\phi}^{2}}{2} + \frac{\dot{\chi}^{2}}{2}
+\frac{V(\phi, \chi)}{f(\phi, \chi)}
 \right)
\end{eqnarray}
and 
\begin{eqnarray}
\dot{H} + H^{2} =
-\frac{1}{3f(\phi, \chi)}
\left( \frac{\dot{\phi}^{2}}{2} + \frac{\dot{\chi}^{2}}{2}
-\frac{V(\phi, \chi)}{f(\phi, \chi)}\right)
\end{eqnarray}
with $H(t)$ being the Hubble parameter.

Field fluctuations will be studied later where one expands each scalar field about its background value such that, to first order in perturbations, one writes $\phi(x) = \phi(t) + \delta \phi(t, \vec{x})$ and $\chi(x) = \chi(t) + \delta \chi(t, \vec{x})$. In this regard, one shows that, unlike the single field case, two distinct sources of nonadiabatic perturbations arise in the case of multiple fields.  While both sources are a generic feature of multifields, one of them seems to result from nonminimal coupling. Thus, at supper horizon scales, nonadiabatic pressure does not vanish
\begin{eqnarray}
\delta \bm{P}_{\text{nad}} \equiv
\delta \bm{P} - \frac{\dot{\bm{P}}}{\dot{\rho}}\delta \rho \neq 0 \quad \text{at} \quad k \ll aH.
\end{eqnarray}
In the present case of two fields, this quantity arises from two parts 
\begin{eqnarray}
\label{source1}
\delta \bm{P}_{\text{nad}}^{\text{multifield}} =
\frac{2 \dot{\phi} \dot{\chi} \left( \dot{\chi} V_{,\phi} - \dot{\phi} V_{,\chi} \right)}{(\dot{\phi}^{2} + \dot{\chi}^{2})f} 
\left(\frac{\delta \phi}{\dot{\phi}} -\frac{\delta \chi}{\dot{\chi}} \right)
\end{eqnarray}
and
\begin{eqnarray}
\label{source2}
\delta \bm{P}_{\text{nad}}^{\text{Nonmin}} =
\frac{4 V \dot{\phi} \dot{\chi} \left( \dot{\chi} f_{,\phi} - \dot{\phi} f_{,\chi} \right)}{(\dot{\phi}^{2} + \dot{\chi}^{2})f^{3}} \left(\frac{\delta \phi}{\dot{\phi}} -\frac{\delta \chi}{\dot{\chi}} \right)
\end{eqnarray}
where comma refers to derivatives with respect to the corresponding field.

Few remarks on these sources of nonadiabaticity are in order:
\begin{itemize}
    \item In the limit where $f=1$, the first source (\ref{source1}) coincides with the familiar nonadiabatic pressure that arises in GR where both fields are coupled minimally to gravity in its metrical form. The reason for this is that purely affine gravity gets reduced to GR in the case of minimal coupling \cite{affine_inflation}.
    \item For the same case, as expected, the second source disappears leaving only one source of entropy perturbation. However, an interesting result here is that the source (\ref{source2}) also vanishes when only one scalar field  is considered, and  thus, the nonminimal coupling contribution to entropy production is valid only in the case of multifields. This result must also be expected since in the single field limit, the theory can always be transformed to a minimal coupling case where the field gains a canonical kinetic term, and hence, the perturbations are adiabatic. This is one of the cases in which the change of frame (Jordan to Einstein) in metrical theories leads to some confusion \cite{sasaki} (see also \cite{frame_issue_in_general} for the frame-issue in general.)    
\end{itemize}
In the subsequent sections we will be interested in the minimal coupling case. We will derive the main equations of motion by developing the covariant formalism in the field space manifold. 

\subsection{Minimal couplings and covariant formalism}

In this section we study the transition to minimal coupling so that the gravitational sector gains a canonical form. At first sight, this will appear to be similar to the transformation from Jordan to Einstein frame in GR, however, we will show that unlike metric theories of gravity, this transition is performed using only potential redefinitions without applying any conformal transformation of the metric.

First of all, let us generalize action (\ref{action1}) by involving a noncanonical fields with a non-Euclidean field space metric $\bm{k}_{ab}(\bm{\phi})$. In this case, the gravitational action is written as
\begin{eqnarray}
\label{action-nonminimal}
S[\Gamma, \bm{\phi}]=
\int d^{4}x \frac{\sqrt{ | f(\bm{\phi}) \bm{R}_{\mu\nu}(\Gamma)- \bm{k}_{ab}(\bm{\phi})\nabla_{\mu}\bm{\phi}^{a} \nabla_{\nu}\bm{\phi}^{b} |}}
{V(\bm{\phi})} \nonumber \\
\end{eqnarray}

As it has been performed in the previous section, one can directly derive the field equations by following the above steps. However, in what follows we will be interested in the transition to minimal coupling where the gravitational sector enjoys a canonical form. In this regard, one can easily show that this action is recast into

\begin{eqnarray}
\label{action-minimal}
S[\Gamma, \bm{\phi}]=
\int d^{4}x \frac{\sqrt{ | M^{2}_{\text{Pl}} \bm{R}_{\mu\nu}(\Gamma)- \bm{\mathcal{G}}_{ab}(\bm{\phi})\nabla_{\mu}\bm{\phi}^{a} \nabla_{\nu}\bm{\phi}^{b} |}}
{U(\bm{\phi})} \nonumber \\
\end{eqnarray}
where the potential is simply redefined as 
\begin{eqnarray}
\label{potential-new}
V(\bm{\phi})\rightarrow U(\bm{\phi})=
\frac{M^{4}_{\text{Pl}}}{f^{2}(\bm{\phi})}V(\bm{\phi}).
\end{eqnarray}
Additionally, the transformation to minimal coupling induces a new curved metric 
\begin{eqnarray}
\label{field_space_metric}
\bm{\mathcal{G}}_{ab}(\bm{\phi})=
\frac{M^{2}_{\text{Pl}}}{f(\bm{\phi})} \bm{k}_{ab}(\bm{\phi}).
\end{eqnarray}
The field space with this metric differs crucially from that obtained in GR with nonminimal coupling. The differences will be discussed below when studying various shapes of the manifold. 

In what follows, we will adopt the covariant formalism to study the primordial perturbations in inflationary dynamics resulting from (\ref{action-minimal}). First, each scalar of the $N$ fields $\bm{\phi}^{a}(x)$ will be expanded around its background as
\begin{eqnarray}
\label{field_fluctuation}
\bm{\phi}^{a}(x) \rightarrow \bm{\phi}^{a}(t) + \delta\bm{\phi}^{a}(x).
\end{eqnarray}
Here, the set of the classical backgrounds $\bm{\phi}^{a}(t)$ describes classical paths in the field space as do the general coordinates in curved manifolds. However, field displacements as well as derivatives of the fields do manifest as vectors in the field space \cite{kaiser}. To that end, one defines a covariant derivative as the following operator applied to any vector, for instance $\delta\bm{\phi}^{a}$, as
\begin{eqnarray}
\bm{D}_{c}\delta\bm{\phi}^{a} = \partial_{c}\delta\bm{\phi}^{a}
+\bm{\Gamma}^{a}_{\,\,cb}\delta\bm{\phi}^{b},
\end{eqnarray}
and a covariant derivative with respect to cosmic time as
\begin{eqnarray}
\bm{D}_{t}\delta\bm{\phi}^{a} = \dot{\bm{\phi}}^{c}\bm{D}_{c}\delta\bm{\phi}^{a},
\end{eqnarray}
where $\bm{\Gamma}^{a}_{\,\,cb}$ are the components of the field space connection, i.e, the Christoffel symbol of the metric (\ref{field_space_metric}). 

The dynamics of the fields $\bm{\phi}^{a}$ is governed by the equations of motion resulting from variation of action (\ref{action-minimal}) with respect to these fields, and read
\begin{eqnarray}
\label{equation_for_phi_minimal}
&&\partial_{\mu}\left(\frac{\sqrt{|\bar{K} (\Gamma, \bm{\phi})|}}{U(\bm{\phi})}(\bar{K}^{-1})^{\mu\nu} 
\partial_{\nu}\bm{\phi}^{b}
\bm{\mathcal{G}}_{ab}(\bm{\phi})\right) \nonumber \\
&&-\frac{1}{2}\frac{\sqrt{|\bar{K}(\Gamma, \bm{\phi})|}}{U(\bm{\phi})}(\bar{K}^{-1})^{\mu\nu} 
\partial_{\mu}\bm{\phi}^{b}\partial_{\nu}\bm{\phi}^{c}
\bm{\mathcal{G}}_{bc,a} \nonumber \\
&&- \frac{\sqrt{|\bar{K}(\Gamma, \bm{\phi})|}}{U^{2}(\bm{\phi})}U_{,a} =0,
\end{eqnarray}
where in this case the tensor field $\bar{K}_{\mu\nu}$ is 
\begin{eqnarray}
\label{Kmin}
\bar{K}_{\mu\nu}(\Gamma, \bm{\phi})=
M^{2}_{\text{Pl}} \bm{R}_{\mu\nu}(\Gamma)- \bm{\mathcal{G}}_{ab}(\bm{\phi})\nabla_{\mu}\bm{\phi}^{a} \nabla_{\nu}\bm{\phi}^{b}.
\end{eqnarray}
Thus, equation (\ref{equation_for_phi_minimal}) involves the spacetime connection and the fields but not the spacetime metric. The latter will be incorporated into the dynamics \textit{a posteriori} after we generate it from the gravitational equation. As in the philosophy of affine gravity \cite{affine_inflation}, the metric is generated dynamically by varying action (\ref{action-minimal}) with respect to the spacetime connection $\Gamma$ which yields
\begin{eqnarray}
\nabla_{\lambda}\left( \frac{\sqrt{|\bar{K}(\Gamma, \bm{\phi})|}}{U(\bm{\phi})}(\bar{K}^{-1})^{\mu\nu} \right)=0,
\end{eqnarray}
and solved as
\begin{eqnarray}
\label{metric_minimal_case}
\frac{\sqrt{|\bar{K}(\Gamma, \bm{\phi})|}}{U(\bm{\phi})}(\bar{K}^{-1})^{\mu\nu}=\sqrt{|g|}(g^{-1})^{\mu\nu}.
\end{eqnarray}
Therefore, following the same steps made in (\ref{equations_in_terms_of_ricci}) to (\ref{eom-gravity}), one finds the gravitational field equations
\begin{eqnarray}
\label{einstein_equations_minimal}
M^{2}_{\text{Pl}}\left(R_{\mu\nu} - \frac{1}{2}g_{\mu\nu}R\right) = && 
\bm{\mathcal{G}}_{ab} \nabla_{\mu}\bm{\phi}^{a} \nabla_{\nu} \bm{\phi}^{b} \nonumber \\
&&-\frac{1}{2}g_{\mu\nu} 
g^{\alpha\beta}\bm{\mathcal{G}}_{ab} \nabla_{\alpha}\bm{\phi}^{a} \nabla_{\beta} \bm{\phi}^{b} \nonumber \\
&&-g_{\mu\nu}U(\bm{\phi})
\end{eqnarray}

We should emphasize that the metric $g_{\mu\nu}$ which is generated here (case of minimal coupling) coincides with the metric tensor generated in the nonminimal coupling case as in (\ref{metric}). This can be seen easily from (\ref{metric_minimal_case}) by switching back to $V(\phi)$ and $K_{\mu\nu}(\Gamma,\bm{\phi})$. In other words, unlike metric and Palatini formulations where the transition from Jordan to Einstein frame necessitates (metric) conformal transformation, here the gravitational field equations (\ref{eom-gravity}) and (\ref{einstein_equations_minimal}) are related to each other by potential transformation only.
The origin of this feature can be understood from the structure of the purely affine action (\ref{action1}) in which the potential energy enters the denominator. This allows for rescaling kinetic and potential energies so that the ratio which manifests as a spacetime metric in both scenarios (minimal and nonminimal) is unique. 

This new feature might have interesting consequences on the cosmological parameters in both minimal and nonminimal coupling theories. The uniqueness of the metric implies that the rate of expansion is the same in both minimal and nonminimal scenarios. Any field and potential redefinition used for the transition between (\ref{eom-gravity}) and (\ref{einstein_equations_minimal}) has no effects on the Hubble parameter which is related only to the spacetime metric (the scale factor.) This shows that the number of e-folds, and more interestingly the gauge invariant curvature perturbation, are not altered by this change between actions. 
Hence, the inflationary dynamics based on purely affine theory is free of any of the confusions caused by changing the frame, as we have discussed above.

In terms of the metric, one finally writes equation (\ref{equation_for_phi_minimal}) in a more familiar form
\begin{eqnarray}
\bm{\mathcal{G}}_{ab}\Box \bm{\phi}^{b} +
\left(\bm{\mathcal{G}}_{ab,c} -\frac{1}{2}\bm{\mathcal{G}}_{bc,a} \right)g^{\mu\nu}\nabla_{\mu}\bm{\phi}^{b} \nabla_{\nu} \bm{\phi}^{c}- U_{,a}=0. \nonumber \\
\end{eqnarray}
With a bit more manipulation, this equation takes finally the standard form  
\begin{eqnarray}
\label{box_phi}
\Box \bm{\phi}^{a} +\bm{\Gamma}^{a}_{\,\,bc}g^{\mu\nu}\nabla_{\mu}\bm{\phi}^{b} \nabla_{\nu} \bm{\phi}^{c} - \bm{\mathcal{G}}^{ab}U_{,b}=0.
\end{eqnarray}
Although this equation, which is derived via the covariant formalism, takes a standard form as in the case of metric gravity, there is a crucial difference, however. The origin of the differences comes from the  manifold's metric (\ref{field_space_metric}). In fact, in the purely metric gravity, the transition to Einstein frame would induce a field space metric of the form \cite{sasaki}
\begin{eqnarray}
\label{field_space_metric_gr}
\bm{\mathcal{G}}_{ab}(\bm{\phi})=
\frac{M^{2}_{\text{Pl}}}{2f(\bm{\phi})}\left(\bm{k}_{ab}(\bm{\phi}) +\frac{3}{f\bm{(\phi})}f_{,a}f_{,b} \right).
\end{eqnarray}

The second term which is proportional to the derivative of the nonminimal coupling function $f(\bm{\phi})$ arises from the nonlinearity (in the metric) of the Ricci scalar. This induces kinetic terms for the scalar fields in Einstein frame. However, it is clear that the metric (\ref{field_space_metric}) is free of these additional terms thanks to the linearity of the spacetime curvature in the affine connection.  This makes the purely affine formulation distinguishable from its metrical counterpart. 

Now, the $N$-field fluctuations (\ref{field_fluctuation}) will be adapted to the first order scalar perturbations of the spacetime metric
\begin{eqnarray}
\label{metric_perturbation}
ds^{2} =&& -(1+A)dt^{2}
+ 2a(t)\partial_{i}Bdx^{i}dt \nonumber \\
&&+a^{2}(t)\left[(1-2\psi)\delta_{ij} + 2\partial_{i}\partial_{j}E \right]dx^{i}dx^{j},
\end{eqnarray}
with standard notations for the scale factor and the scalar gauge degrees of freedom.

The feature of a uniquely generated metric which we have discussed above shows that the perturbed metric (\ref{metric_perturbation}) is the same as that one uses in the minimal coupling case.   

The background part of the dynamics arises from Einstein field equations as
\begin{eqnarray}
\label{hubble_background}
&&H^{2} = \frac{1}{3M^{2}_{\text{Pl}}}\left(\frac{1}{2}\bm{\mathcal{G}}_{ab}\dot{\bm{\phi}}^{a}(t)\dot{\bm{\phi}}^{b}(t) +U(\bm{\phi})\right) \\
&& \dot{H}= -\frac{1}{2M^{2}_{\text{Pl}}}\bm{\mathcal{G}}_{ab}\dot{\bm{\phi}}^{a}(t)\dot{\bm{\phi}}^{b}(t).
\end{eqnarray}
For the perturbation part, one introduces the familiar Mukhanov-Sasaki quantity
\begin{eqnarray}
Q^{a} = \mathcal{Q}^{a}
+\frac{\dot{\bm{\phi}}}{H}\psi,
\end{eqnarray}
such that the vector $\mathcal{Q}^{a}$, which is introduced for the purpose of the covariant representation of field fluctuations, coincides with $\delta \bm{\phi}^{a}$ at first order in perturbation \cite{covariant_approach}.

With the covariant approach, equation (\ref{box_phi}) splits into background and perturbations parts where the former reads
\begin{eqnarray}
\label{background_evol}
\bm{D}_{t}\dot{\bm{\phi}}^{a}
+3 H \dot{\bm{\phi}}^{a}
+\bm{\mathcal{G}}^{ab}U_{,b} =0.
\end{eqnarray}
The perturbed part of the equation takes the form \cite{kaiser_covariant}
\begin{eqnarray}
\label{eq_for_fluctuation}
&&\bm{D}^{2}_{t} Q^{a} + 3H\bm{D}_{t} Q^{a} \nonumber \\ 
&&+\Bigg\{\frac{k^{2}}{a^{2}}\delta^{a}_{\,b} + \mathcal{M}^{a}_{\,b} 
- \frac{1}{M^{2}_{\text{Pl}}a^{3}} \bm{D}_{t}\left(\frac{a^{3}}{H}\dot{\bm{\phi}}^{a}\dot{\bm{\phi}}_{b} \right) \Bigg\} Q^{b}=0. \nonumber \\
\end{eqnarray}
Here, the matrix $\mathcal{M}^{a}_{\,b}$, playing the role of an effective mass-square, is written in terms of the Riemann tensor $\mathfrak{R}^{a}_{\,bcd}$ of the curved (field space) manifold as
\begin{eqnarray}
\mathcal{M}^{a}_{\,b}=
\bm{\mathcal{G}}^{ac}\bm{D}_{b}\bm{D}_{c}U(\bm{\phi})
- \mathfrak{R}^{a}_{\,cdb}\dot{\bm{\phi}}^{c}\dot{\bm{\phi}}^{d}.
\end{eqnarray}
It is also useful to introduce the quantity $\dot{\sigma}^{2}= \bm{\mathcal{G}}_{ab}\dot{\bm{\phi}}^{a}\dot{\bm{\phi}}^{b}$, the magnitude of the background fields' vector, and the unit vector $\hat{\sigma}^{a} = \dot{\phi}^{a}/\dot{\sigma}$. Another important quantities are, the turn-rate of the background fields $\omega^{a}= \bm{D}_{t}\sigma^{a}$ with magnitude $\omega =|\omega^{a}|$ and the vector $\hat{s}= \omega^{a}= \omega^{a}/\omega$ directed perpendicular to the motion of the fields. With these quantities one is able to decompose the vector fluctuations into adiabatic and entropic perturbations given by $Q_{\sigma}=\hat{\sigma}_{a}Q^{a}$ and $Q_{s}=\hat{s}_{a}Q^{a}$ respectively.  

The previous equation splits into two parts describing the adiabatic
\begin{eqnarray}
&&\ddot{Q}_{\sigma} + 3H \dot{Q}_{\sigma}  
+\Bigg\{\frac{k^{2}}{a^{2}} + \mathcal{M}_{\sigma \sigma} 
-\omega^{2}- \frac{a^{-3}}{M^{2}_{\text{Pl}}} \frac{d}{dt}\left(\frac{a^{3}}{H}\dot{\sigma}^{2} \right) \Bigg\} Q_{\sigma} \nonumber \\
&&=2\frac{d}{dt}(\omega Q_{s}) -2\left(\frac{V_{,\sigma}}{\dot{\sigma}} + \frac{\dot{H}}{H} \right)
\omega Q_{s} 
\end{eqnarray}
and isocurvature    
\begin{eqnarray}
\ddot{Q}_{s} + 3H \dot{Q}_{s}  
+\Bigg\{\frac{k^{2}}{a^{2}} + \mathcal{M}_{ss} 
+3\omega^{2} \Bigg\} Q_{s} 
= 4 M^{2}_{\text{Pl}}\frac{\omega}{\dot{\sigma}}\frac{k^{2}}{a^{2}}\Psi,
\nonumber \\
\end{eqnarray}
modes, where $\Psi$ is the Bardeen potential and 
\begin{eqnarray}
\mathcal{M}_{\sigma\sigma} = \hat{\sigma}_{a}\hat{\sigma}^{b}\mathcal{M}^{a}_{b},
\quad
\mathcal{M}_{ss} = \hat{s}_{a}\hat{s}^{b}\mathcal{M}^{a}_{b}.
\end{eqnarray}

In what follows we will describe some interesting shapes of the field manifold. 

\subsection{Conformally flat field space}

An interesting case is when both fields gain canonical kinetic terms, i.e, $\bm{k}_{ab}(\bm{\phi}) \rightarrow \bm{\delta}_{ab}$. Thus, unlike metric gravity (see equation (\ref{field_space_metric_gr})), the field space metric (\ref{field_space_metric}) which reads
\begin{eqnarray}
\label{conformal_field_space_metric}
\bm{\mathcal{G}}_{ab}(\bm{\phi})=
\frac{M^{2}_{\text{Pl}}}{f(\bm{\phi})} \bm{\delta}_{ab}.
\end{eqnarray}
is conformally flat. 

Needless to say, this property of conformally flat space is not restricted to two-dimensional field space; it occurs in general dimensions as long as the fields with their nonminimal couplings are placed in the affine action \cite{entropy_production_in_affine_inflation}.

The nonzero components of the connection of the two-dimensional field space, with coordinates $\bm{\phi}^{a}=(\phi, \chi)$, can be easily calculated from its metric (\ref{field_space_metric}) and read
\begin{eqnarray}
\bm{\Gamma}^{\phi}_{\,\,\phi\phi}= -\frac{f_{,\phi}}{2f}, \quad \bm{\Gamma}^{\phi}_{\,\,\phi\chi}=\bm{\Gamma}^{\phi}_{\,\,\chi\phi}= -\frac{f_{,\chi}}{2f}, \quad \bm{\Gamma}^{\phi}_{\,\,\chi\chi}= \frac{f_{,\phi}}{2f} \nonumber \\
\end{eqnarray}
and by symmetry
\begin{eqnarray}
\bm{\Gamma}^{\chi}_{\,\,\chi\chi}= -\frac{f_{,\chi}}{2f}, \quad \bm{\Gamma}^{\chi}_{\,\,\phi\chi}=\bm{\Gamma}^{\chi}_{\,\,\chi\phi}= -\frac{f_{,\phi}}{2f}, \quad
\bm{\Gamma}^{\chi}_{\,\,\phi\phi}= \frac{f_{,\chi}}{2f}
\nonumber \\
\end{eqnarray}

While for $N > 2$ dimensions the flatness of the field space necessitates the vanishing of its Riemann tensor, in two dimensions it is sufficient to have a zero Ricci scalar for the field space to be totally Euclidean. In the latter case, one can easily show that the Ricci scalar of the induced metric (\ref{field_space_metric}) with space coordinates $\phi$ and $\chi$, takes the form     
\begin{eqnarray}
\label{curvature_of_field_space}
\mathfrak{R}(\bm{\mathcal{G}}) =
\frac{1}{M^{2}_{\text{Pl}}}\Bigg\{\frac{\partial^{2}f}{\partial \phi^{2}}
+\frac{\partial^{2}f}{\partial \chi^{2}}
- \frac{1}{f} \left(\frac{\partial f}{\partial \phi} \right)^{2}
- \frac{1}{f} \left(\frac{\partial f}{\partial \chi} \right)^{2} \Bigg\} \nonumber \\
\end{eqnarray}
This expression of the scalar curvature shows that, in general, it is not allowed to bring the field space metric (\ref{field_space_metric}) to the  Euclidean form. In other words, no field transformations for a general coupling function $f(\phi, \chi)$ can recast the kinetic terms of both scalar fields into canonical forms simultaneously. Nevertheless, possible choices of the nonminimal coupling function may render the above curvature to zero for specific constraints on the coupling parameters.

\subsection{Flattening the field space and inflationary dynamics}
As we have seen so far, the field space cannot be made flat since there is no field transformation according to which the metric (\ref{field_space_metric}) is Euclidean. However, a more interesting feature that cannot be realized in metric gravity is that one is able to flatten the field space by applying equal couplings for the spacetime curvature and the kinetic terms of the fields. This is performed in action (\ref{action-nonminimal}) by letting $\bm{k}_{ab}(\bm{\phi})=(f(\bm{\phi})/M^{2}_{\text{Pl}})\bm{\delta}_{ab}$, and therefore the metric (\ref{field_space_metric}) reads $\mathcal{G}_{ab}=\bm{\delta}_{ab}$ while the potential keeps the same redefinition (\ref{potential-new}).

\begin{figure}[!t]
\includegraphics[width=\columnwidth]{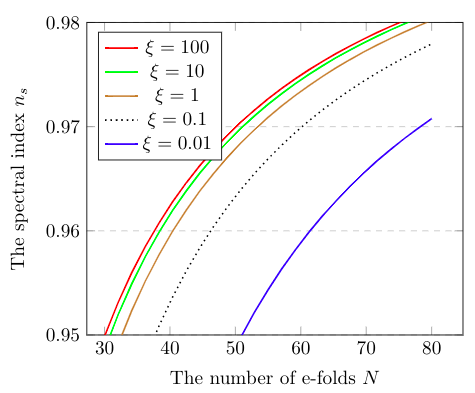}
\caption{The scalar tilt in terms of the number of e-folds that measure the duration of inflation. One notices that larger values of $\xi$ require a small number of e-folds for the spectral index to fall in the observed range. Small $\xi$ is compatible with the observed $n_{s}$ for the usual range $N=50-60$.}
\label{fig: first order spectral index}
\end{figure}

\subsubsection{Single field limit}
In this case, a single field $\phi$ with nonminimal coupling function
$f(\phi) = M^{2}_{\text{Pl}} + \xi_{\phi}\phi^{2}$, a non-canonical kinetic coupling term
$\bm{k}(\phi) = f(\phi)/M^{2}_{\text{Pl}}$ and a potential $V(\phi) = (\lambda_{\phi}/4)\phi^{4}$, will be characterized by the following slow-roll dynamics
\begin{eqnarray}
\label{slow_roll_single}
3M^{2}_{\text{Pl}}H^{2} \simeq U(\bm{\phi}), \quad
3H\dot{\phi} \simeq - U^{\prime}(\phi),
\end{eqnarray}
where the new potential is given by (\ref{potential-new}) for the single field limit, i.e, 
\begin{eqnarray}
\label{potential_single}
U(\phi) = \frac{M^{4}_{\text{Pl}}\lambda_{\phi} \phi^{4}}{4(M^{2}_{\text{Pl}} + \xi_{\phi}\phi^{2})^{2}}
\end{eqnarray}

The model described by (\ref{slow_roll_single})-(\ref{potential_single}) differs significantly from the familiar single field quartic models with nonminimal coupling. The novelty here is that the quartic potential becomes flat (assuming large fields) due to the nonminimal coupling function $f(\phi)$ without altering the canonical kinetic terms thanks to  $\bm{k}(\bm{\phi})=f(\bm{\phi})/M^{2}_{\text{Pl}}$. Thus, the inflationary dynamics described mainly by the slow-roll equations takes a simple form even in the case of nonminimal interaction. In fact, one calculates the main inflationary parameters such as the number of e-folds and slow-roll parameters using only the original field $\phi$ without the aid of an auxiliary field that usually describes the inflaton in Einstein frame. Therefore, one expects new predictions compared to that obtained from the same model but in the GR framework.     

This allows one to easily obtain the number of e-foldings $N$ that measures the duration of inflation as $N \simeq \xi_{\phi}\phi^{4}/16$,
%\begin{eqnarray}
%N = \int_{\phi_{\text{end}}}^{\phi_{*}}
%\frac{U(\phi)}{U^{\prime}(\phi)}d \phi \simeq
%\end{eqnarray}
leading to the field's value at the horizon crossing $\phi_{*}/M_{\text{Pl}} \simeq (16 N_{*}/\xi_{\phi})^{1/4}$. The first order slow-roll parameters at this value read
\begin{eqnarray}
\epsilon
%\frac{1}{2}\left(\frac{U^{\prime}}{U} \right)^{2}
\simeq \frac{8}{(16N_{*})^{3/2}\sqrt{\xi_{\phi}}} \quad \text{and} \quad
\eta
%\frac{U^{\prime \prime}}{U}
\simeq -\frac{3}{4N_{*}}.  
%\nonumber \\
\end{eqnarray}
Therefore, the amplitude of the observed scalar density perturbation is obtained as 
\begin{eqnarray}
%\overset{\mathrm{CMB}}{=}
\bm{A}_{s} \equiv && \frac{U_{*}}{24\pi^{2}\epsilon_{*}}
\xrightarrow{\text{CMB}} 2.1 \times 10^{-9} \nonumber \\
&&\simeq 8 \times 10^{-3} \lambda_{\phi} (N_{*}/\xi_{\phi})^{3/2},
\end{eqnarray}
which implies
\begin{eqnarray}
\label{xi from amplitude}
\xi_{\phi} \simeq 2.5 \times 10^{4} \lambda_{\phi}^{2/3}N_{*}.
\end{eqnarray}

\begin{figure*}[!t]
\includegraphics[width=\columnwidth]{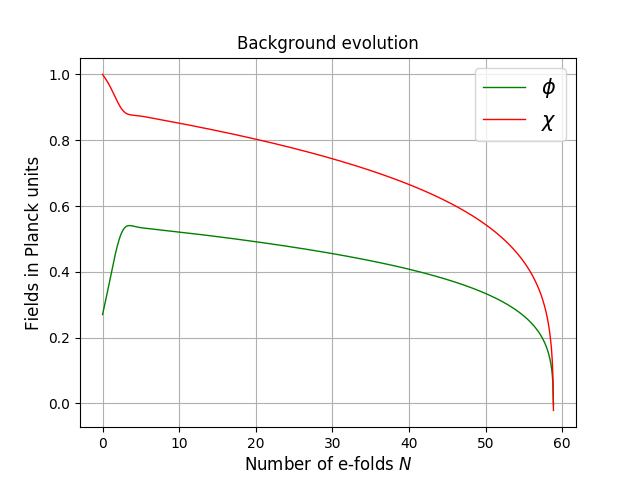}
\includegraphics[width=\columnwidth]{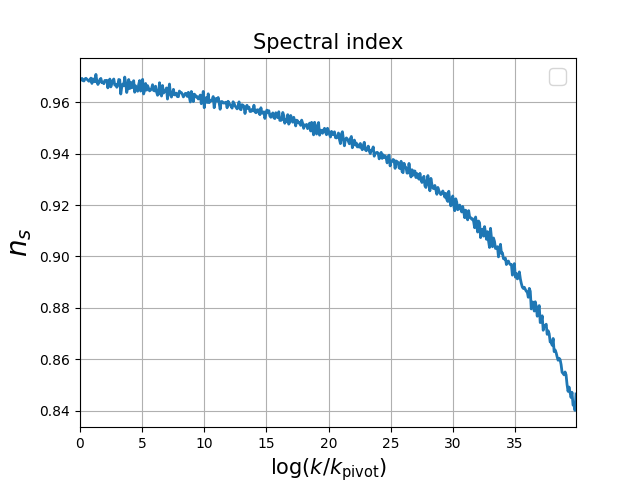}
\caption{Left: Evolution of the background fields in terms of the number of e-folds. This represents the numerical solution of the slow-roll equations (\ref{two_field_hubble})-(\ref{chi_equation}) for initial values $\phi = 0.27$ and $\chi = 1$ (in $M_{\text{Pl}}$), potential parameters $\lambda_{\phi} =2.4 \times 10^{-3}$, $\lambda_{\chi} =3\times 10^{-2}$ and $\lambda_{\text{int}} =2 \times 10^{-4}$. The nonminimal coupling paramters are taken $\xi_{\phi} = 1.8 \times 10^{5}$ and $\xi_{\chi} = 10^{3}$. Right: The numerical solution for the spectral index for the same field's initial values and model parameters used in the background evolution.}
\label{fig:back_ns}
\end{figure*}

Another important observable quantities in inflation are the spectral index $n_{s}$ and the tensor-to-scalar ratio $r$ which can be calculated in this model from the above slow-roll parameters and read
\begin{eqnarray}
\label{ns-r}
n_{s} \simeq 1 - \frac{3}{2N_{*}} - \frac{48}{(16N_{*})^{3/2}\sqrt{\xi_{\phi}}}, \quad
r \simeq \frac{128}{(16N_{*})^{3/2}\sqrt{\xi_{\phi}}}. \nonumber \\
\end{eqnarray}

According to (\ref{xi from amplitude}), we notice that for the SM-Higgs where $\lambda_{h} \sim \mathcal{O}(1)$, the amount of density perturbation would require a large parameter $\xi_{h} \simeq 10^{4} N_{*}$. In this case, the scalar tilt in (\ref{ns-r}) will be dominated by their first two terms (see figure \ref{fig: first order spectral index}), and therefore, the Planck result $n_{s} \simeq 0.965$ is attained here for an amount of inflation, $N<50$. Here, one can easily check that the large values of $\xi_{\phi}$ produce a very small amount for the tensor-to-scalar ratio, namely, $r \sim 10^{-6}$. It is worth mentioning that this ratio is larger than the tiny value of the familiar Palatini-Higgs inflation \cite{palatini_higgs_inflation} or previous single-field affine inflation \cite{affine_inflation}. Current CMB measurements allow only for an upper bound, $r<0.06$, and future observations are expected to proclaim more accurate data for the primordial tensor perturbations. 

\subsubsection{Two-field model and non-Gaussianity}
Here, we consider a potential that manifests in powers of the fields up to quartic terms
\begin{eqnarray}
\label{original quartic potential}
V(\phi,\chi) = \frac{\lambda_{\phi}}{4}\phi^{4}
+\frac{\lambda_{\text{int}}}{2}\phi^{2}\chi^{2}
+ \frac{\lambda_{\chi}}{4}\chi^{4},
\end{eqnarray}
where $\lambda_{\phi}$, $\lambda_{\text{int}}$ and $\lambda_{\text{int}}$ are dimensionless coupling constants.

With a nonzero dimensionless couplings $\xi_{\phi}, \xi_{\chi}$, a relevant nonminimal coupling function takes the form
\begin{eqnarray}
f(\phi, \chi)=
M^{2}_{\text{Pl}} + \xi_{\phi}\phi^{2} + \xi_{\chi}\chi^{2}, 
\end{eqnarray}
leading to a potential (\ref{potential-new})
\begin{eqnarray}
\label{two_field_new_potential}
U(\phi, \chi)= \frac{ M^{4}_{\text{Pl}}}{4}
\frac{(\lambda_{\phi}\phi^{4} +2\lambda_{\text{int}}\phi^{2}\chi^{2}
+ \lambda_{\chi}\chi^{4})  }{(M^{2}_{\text{Pl}} + \xi_{\phi}\phi^{2} + \xi_{\chi}\chi^{2})^{2}}.
\end{eqnarray}

The single filed-limit of this model is studied above (see (\ref{slow_roll_single})-(\ref{potential_single})).

We notice that there is no symmetry that forces the coupling constants $\lambda_{a}$ and $\xi_{a}$ (with $a = \phi,\chi$) to gain the same value. This means that, in general, this potential does not obey a rotational symmetry, the fact that makes it difficult to write it totally in terms of a single radial field. In the strong field regime, the flatness of the potential can be seen in a chosen direction (for a large field component $\bm{\phi}^{a}=\phi$ (or $\chi$)) as 
\begin{eqnarray}
U(\bm{\phi}^{a}) \simeq
\frac{M^{4}_{\text{Pl}} \lambda_{a}}{4\xi^{2}}
\left[1 + \mathcal{O}\left( \frac{M^{2}_{\text{Pl}}}{\xi (\bm{\phi}^{a})^{2}}\right) \right]
\end{eqnarray}
without summation convention.

The background evolution $\phi(t)$ and $\chi(t)$ is now described by their equations of motion (\ref{hubble_background}) and (\ref{background_evol}). In this case where the field space metric is reduced to the flat metric, hence, the connection coefficients vanish, one obtains the coupled equations  
\begin{eqnarray}
\label{two_field_hubble}
&&3M^{2}_{\text{Pl}}H^{2} = \frac{1}{2}\dot{\phi}^{2}(t)
+\frac{1}{2}\dot{\chi}^{2}(t) +U(\bm{\phi}), \\
&&\ddot{\phi} + 3H\dot{\phi} + \frac{\partial U}{\partial \phi}=0,
\label{phi_equation}
\quad
\ddot{\chi} + 3H\dot{\chi} + \frac{\partial U}{\partial \chi}=0
\label{chi_equation}
\end{eqnarray}

Notice again that unlike the familiar models with nonminimal couplings to gravity which involve, in Einstein frame, both a new (or redefined) potential and non-canonical kinetic terms, in the present model we have only a redefined potential. 

\begin{figure*}[!t]
%\centering
\includegraphics[width=0.49\textwidth]{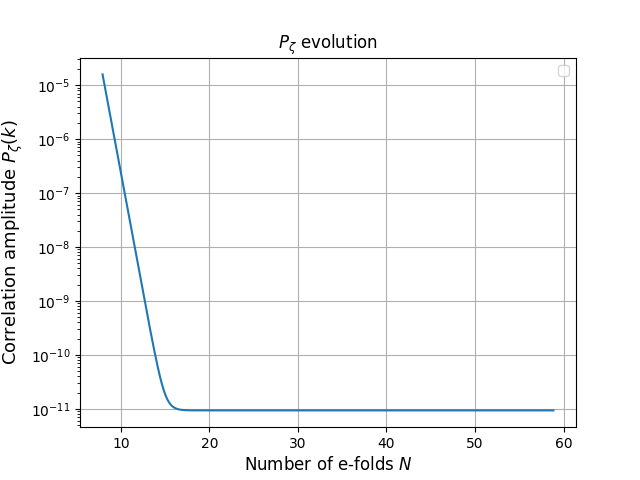}
\includegraphics[width=0.49\textwidth]{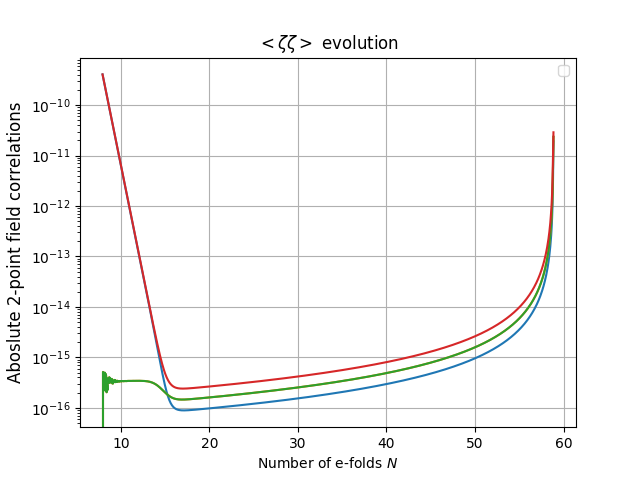}\\
\includegraphics[width=0.49\textwidth]{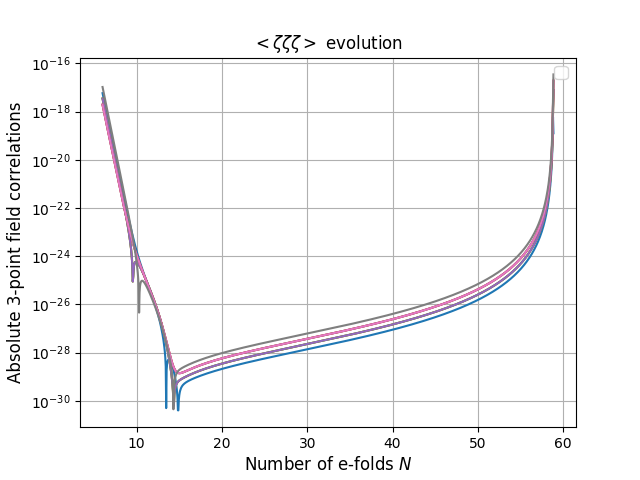}
\includegraphics[width=0.49\textwidth]{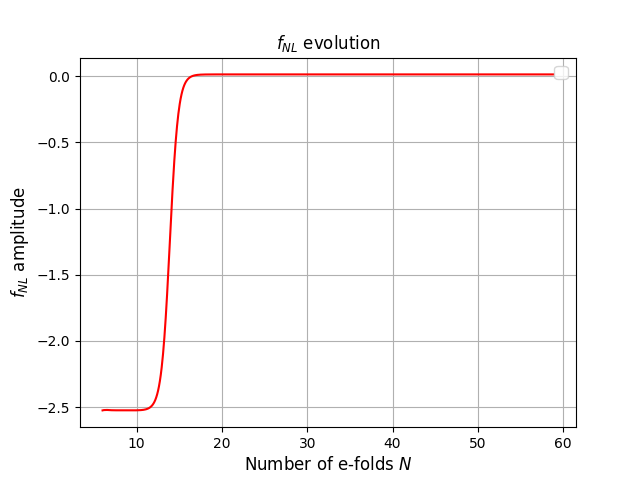}
\caption{Up: The power spectrum of the curvature perturbation $\zeta$ and the two-point correlation function. Down: The three-point function and the reduced bi-spectrum $f_{NL}$. The numerical analysis are performed using
the couplings $\lambda_{\phi}=2.4 \times 10^{-3}$, $\lambda_{\text{int}}=2 \times 10^{-4}$ , and $\lambda_{\chi}=3 \times 10^{-2}$ for the potential (\ref{two_field_new_potential}), and the nonminimal coupling parameters $\xi_{\phi}=1.8 \times 10^{5}$, $\xi_{\chi}= 10^{3}$. For the shape of $f_{NL}$, one notices that the distribution can be less non-Gaussian only for small e-folds.}
\label{fig:all_fig}
\end{figure*}

%The main difficulty one usually encounters when studying two field inflation lies in the fields' initial conditions. In the single field case where the field obeys the slow-roll equation, the initial value of the field, satisfying $N$ e-folds of inflation, can be easily obtained analytically by inverting the equation with respect to time. This is no longer possible when a second (or more) scalar enters the dynamics, since in this case more than one solution is possible. In fact, the slow-roll equation that arises from (\ref{box_phi}) shows that there is a non trivial two-dimensional zone containing the initial values of the two fields that satisfy the required number of e-folds. Generally speaking, there is no analytical manner that can lead to background solutions.

The last coupled equations as well as the evolution equations (\ref{eq_for_fluctuation}) for the fluctuations (necessary for calculating the correlation functions) cannot be solved analytically. For that,  we have employed the open-source PyTransport code \cite{pytransport_code} to (numerically) compute our predictions (see also \cite{cpp_code} for a related code). The code stands on the $\delta N$ formalism and uses the transport method \cite{transport_method1, transport_method2}. After solving for the background dynamics, in the transport method, one evolves both the two and three-point correlation functions of the field fluctuations on sub-horizon scales, and involves all the tree-level contributions. The correlations are then used to compute the power spectrum and the bi-spectrum \cite{transport_method1, transport_method2}. 

%The coupled equations (\ref{two_field_hubble})-(\ref{chi_equation}) are integrated numerically with initial values for the fields (in $M_{Pl}$), namely $\phi_{0}=0.27$ and $\chi_{0}=1$. 

The evolution of the fields in terms of the number of e-folds $N$, and the spectral index of this affine gravity' two-field model are illustrated in figure (\ref{fig:back_ns}). Here the potential's coupling parameters $\lambda_{a}$ as well as the nonminimal coupling parameters $\xi_{a}$ are chosen to give an estimate for the spectral index that falls into the measured bounds; $n_{s} \simeq 0.965$. With these field parameters, the model provides an amount $r \simeq 2.5 \times 10^{-4}$ for the tensor-to-scalar ratio.

For the time being, the temperature distribution of the CMB tends to be close to a Gaussian distribution. In figure \ref{fig:all_fig} (the upper plots) we depict the evolution of the power spectrum and the correlation function with respect to the amount of inflation $N$. These numerical results are found using the same parameter constants, of the potential and the nonminimal coupling function, that we have utilized for the previous background evolution. Power spectrum (or the two-point function) must contain all the characteristics of the perturbations.

Another interesting feature of two-field inflation (and multifields in general) is its prediction for a non-Gaussian distribution for the primordial perturbation. Deviations from Gaussianity of the primordial perturbations can be tracked via the possible nonzero three-point correlation function (or bi-spectrum). The evolution of the three-point function as well as the reduced bi-spectrum $f_{NL}$ are given in figure \ref{fig:all_fig} (lower plots). The results show that the distribution can be less Gaussian only for small e-folds. For larger e-folds, $f_{NL} \sim 0$ and the distribution cannot be distinguishable from the Gaussian one. Recent Planck results provides probing value $f_{NL}^{\text{local}} \simeq -0.9 \pm 5.1$ using combined temperature and polarization analysis as well as low-multipole ($4 \leq \ell < 40$) polarization data \cite{planck}. Reducing the $f_{NL}$ errors and improving this constraint will be one of the next challenges for future cosmological observations.     

We conclude by stating few points concerning the two-field Higgs inflation in this framework. Here, the above model could be improved by an $SU(2)$ gauge symmetry which then will require an equal minimal coupling parameters $\xi_{\phi}=\xi_{\chi}=\xi$ for both scalar fields, and the potential will read
\begin{eqnarray}
\label{higgs potential einstein frame}
U\left(h, \chi \right)=
\frac{\lambda M^{4}_{\text{Pl}}\left(h^{2} + \chi^{2} - v^{2} \right)^{2} }
{4 \left(M^{2}_{\text{Pl}}+ \xi \left(h^{2}
+ \chi^{2}\right) \right)^{2}}.
\end{eqnarray}
including a standard model scalar Higgs boson $h$, a single Goldstone mode $\chi$ and the vacuum expectation value $v = 246$ GeV. Hence, the model would obey a rotational symmetry $SO(2)$ preserving the field's magnitude $\rho=(h^{2} + \chi^{2})^{1/2}$. With the SM self coupling $\lambda \sim \mathcal{O}(1)$, the amplitude of the denisty perturbation will certainly require a strong parameter $\xi$ which shifts the predicted value of the scalar tilt $n_{s}$ from the observed value. Two-field Higgs inflation may require a curved field space in which $\bm{\mathcal{G}}_{ab} \neq \delta_{ab}$. This can be realized in our setup by using $\bm{k}_{ab} = \delta_{ab}$, i.e, when both fields are taken canonical while coupled nonminimally to affine gravity as in action (\ref{action1}). However, unlike the flat trajectories, the curved field space metric usually affects the slow-roll assumption and leads finally to a nonlinear equation containing both, field $\phi(t)$ (or $\chi(t)$) and velocity $\dot{\phi}(t)$ (or $\dot{\chi}(t)$). Despite being challenging even numerically, investigating these solutions in the curved field space is interesting in its own \cite{progress}. 
%\cleardoublepage
\section{Conclusion}
\label{sec:conclusion}
  
In this paper we have studied the inflationary dynamics of multiple fields (two fields in particular), with nonadiabatic perturbations and non-Gaussianity features, by considering the purely affine theory of gravity. It is clear that the strong foundation of relativistic gravity relies on the spacetime curvature as an aspect of the gravitational phenomenon. In principle, this favors introducing spacetime affine connection at the first place instead of metric. 

We have considered scalar fields coupled to the affine curvature which is given solely in terms of the connection. This linear interaction has facilitated the emergence of the metric tensor through the equations of motion. We have shown that when the fields become minimally coupled, the covariant formalism can also be used in this context. We have then stated and discussed two main differences between this formulation and the standard metrical gravity case as follows. (\textit{i}) We have shown that the transformation to the minimal couplings is merely performed without a conformal transformation of the metric. The metric here is kept unchanged while redefining only the potential. This new feature, not found in purely metrical and Palatini formulations, offers a possible way out to the frame-issue (sometimes considered as a true ambiguity.) As we have noticed in this paper, as a consequence this feature, the notion of adiabaticity is invariant under the transition to minimal coupling dynamics. (\textit{ii}) It turned out that, when the fields are canonical the linearity of the curvature in the connection leads to simple conformally flat field space metric which is given in terms of the nonminimal coupling function only. (\textit{iii}) An other interesting feature found in this formalism but not in GR is that the overall field space can be made flat by imposing equal interactions for the nonminimal coupling and the non-canonical kinetric terms, i.e, $\bm{k}_{ab}(\bm{\phi}) \propto f(\bm{\phi})\bm{\delta}_{ab}$, the case where the fields become canonical but the potential get redefined. This simplifies the inflationary dynamics significantly while keeping the effects of the nonminimal coupling function on the potential. 

As an application, we restricted the study to the two-field dynamics where we have focused on the potential with powers of the fields up to quartic terms. We have studied inflation by flattening the field manifold and have shown that even the single field-limit leads to observational predictions that differs significantly from its GR counter part. The quartic potential produces a very small amount of tensor-to-scalar ratio of the order $r \sim 10^{-6}$ for strong coupling to the curvature. This value falls between the predictions of Higgs inflation in its purely metric (GR) and Palatini formulations \cite{higgs_inflation, palatini_higgs_inflation}. We have gone further and performed a numerical solution for the two-field dynamics where we have chosen initial conditions for the fields as well as potential parameters that drives the scalar tilt of the perturbation into the range provided by the observations. With the given parameters we have shown that the ratio can increase to $r \sim 10^{-4}$ though still very small compared to the upper-limit $r < 0.06$ of the CMB measurements. On the other hand, in addition to the power spectrum (the two-point correlation), we have solved for the three-point correlation function numerically and tracked the possible deviation from non-Gaussianity through the reduced bi-spectrum. 

It is worth noting that in the present framework, we have considered only the symmetric part of the spacetime Ricci tensor, a case which is sufficient in describing the gravitational theory. However, inflation in the context of purely affine gravity is expected to have more novel features if the symmetric character of the spacetime Ricci curvature is relaxed \cite{asymmetric}.

\section*{Acknowledgments}
The work is supported by the United Arab Emirates University under UPAR Grant No. 12S004. The authors are grateful to John Ronayne for a fruitful correspondence.
%\cleardoublepage

\end{document}